\documentclass[aps,prl,superscriptaddress,amsfonts,amsmath,amssymb,showpacs,floatfix,preprint,longbibliography]{revtex4-2}

\usepackage{url}
\usepackage{bm}
\usepackage{graphicx}
\usepackage{amsmath}
\usepackage{amstext}
\usepackage{amssymb}
\usepackage{amsfonts}
\usepackage{amsbsy}
\usepackage{verbatim}
\usepackage{xcolor}
\usepackage[colorlinks=true, urlcolor=blue, linkcolor=blue, citecolor=blue, pdftex]{hyperref}
\usepackage{multirow}
\usepackage{floatrow}
\usepackage{float}
\usepackage{gensymb}
\usepackage{textcomp}
\usepackage{enumitem}
\usepackage[version=4]{mhchem}
\usepackage{afterpage}
\usepackage{pbox}
\usepackage{makecell}
\usepackage{array,booktabs}
\usepackage{dsfont}
\usepackage{siunitx}
\usepackage[utf8]{inputenc}
\usepackage{braket}
\usepackage[normalem]{ulem}

\newcommand{\css}{Co$_3$Sn$_2$S$_2$}

\begin{document}


\title{Unravelling the origin of the peculiar transition in the magnetically ordered phase of the Weyl semimetal \css}

\author{Ivica {\v Z}ivkovi{\' c}*}
\email{ivica.zivkovic@epfl.ch}
\affiliation{Laboratory for Quantum Magnetism, Institute of Physics, \'Ecole Polytechnique F\'ed\'erale de Lausanne, CH-1015 Lausanne, Switzerland}

\author{Ravi Yadav}
\affiliation{Institute of Physics, \'Ecole Polytechnique F\'ed\'erale de Lausanne, CH-1015 Lausanne, Switzerland}

\author{Jian-Rui Soh}
\affiliation{Laboratory for Quantum Magnetism, Institute of Physics, \'Ecole Polytechnique F\'ed\'erale de Lausanne, CH-1015 Lausanne, Switzerland}

\author{ChangJiang Yi}
\affiliation{Department of Solid State Chemistry, Max Planck Institute for Chemical Physics of Solids, Dresden 01187, Germany}
\affiliation{Beijing National Laboratory for Condensed Matter Physics, Institute of Physics, Chinese Academy of Sciences, Beijing 100190, China}

\author{YouGuo Shi}
\affiliation{Beijing National Laboratory for Condensed Matter Physics, Institute of Physics, Chinese Academy of Sciences, Beijing 100190, China}
\affiliation{School of Physical Sciences, University of Chinese Academy of Sciences, Beijing 100190, China}
\affiliation{Songshan Lake Materials Laboratory, Dongguan 523808, China}

\author{Oleg V. Yazyev}
\affiliation{Institute of Physics, \'Ecole Polytechnique F\'ed\'erale de Lausanne, CH-1015 Lausanne, Switzerland}

\author{Henrik M. R\o{}nnow}
\affiliation{Laboratory for Quantum Magnetism, Institute of Physics, \'Ecole Polytechnique F\'ed\'erale de Lausanne, CH-1015 Lausanne, Switzerland}

\date{\today}

\begin{abstract}
{Recent discovery of topologically non-trivial behavior in Co$_3$Sn$_2$S$_2$ stimulated a notable interest in this itinerant ferromagnet ($T_\mathrm{C} = 174$\,K). The exact magnetic state remains ambiguous, with several reports indicating the existence of a second transition in the range 125 -- 130\,K, with antiferromagnetic and glassy phases proposed to coexist with the ferromagnetic phase. Using detailed angle-dependent DC and AC magnetization measurements on large, high-quality single crystals we reveal a highly anisotropic behavior of both static and dynamic response of Co$_3$Sn$_2$S$_2$. It is established that many observations related to sharp magnetization changes when $B \parallel c$ are influenced by the demagnetization factor of a sample. On the other hand, a genuine transition has been found at $T_\mathrm{P} = 128$\,K, with the magnetic response being strictly perpendicular to the \textit{c}-axis and several orders of magnitude smaller than for $B \parallel c$. Calculations using density-functional theory indicate that the ground state magnetic structure consist of magnetic moments canted away from the \textit{c}-axis by a small angle ($\sim 1.5^\circ$). We argue that the second transition originates from a small additional canting of moments within the kagome plane, with two equivalent orientations for each spin.}
    
\end{abstract}

\maketitle

Topology and topological properties of matter have recently gained a lot of attention due to the realization of their importance in various exotic phases like topological insulators~\cite{Yan2012}, Dirac and Weyl (semi)metals~\cite{Armitage2018} and spin-liquids~\cite{Savary2017}. The band structure of topologically non-trivial compounds is strongly influenced by spin-orbit coupling (SOC), leading to band-inversion and relativistic fermions with linear dispersions. Depending on whether the spatial ($\mathcal{P}$) and time-reversal ($\mathcal{T}$) symmetries are preserved or broken, the crossing points of such inverted bands are called Dirac or Weyl nodes, respectively. If Weyl nodes are found close to the Fermi energy they can strongly affect the transport properties due to the fact that the nodes act as sources of Berry curvature~\cite{Nagaosa2010}. The control of topological properties using external parameters is highly sought-after, offering a novel type of topological phase transitions. In systems with broken $\mathcal{P}$ this is hard to realize, as they are intrinsically linked to the underlying crystal structure. On the other hand, $\mathcal{T}$-breaking is related to the appearance of magnetic order, allowing temperature $T$ or magnetic field $B$ to tune topological invariants and drive the system across a topological phase transition.

It has been recently suggested that \css\,is a Weyl semimetal~\cite{Liu2018,Liu2019}, exhibiting ferromagnetic (FM) order below $T_\mathrm{C} = 174$\,K~\cite{Schnelle2013}. Evidence of topologically non-trivial behavior includes the anomalous Hall effect~\cite{Liu2018}, the visualisation of surface Fermi arcs~\cite{Liu2019,Morali2019} and the giant magneto-optical response~\cite{Okamura2020}. These are considered to arise from the existence of Weyl nodes with opposite chirality~\cite{Armitage2018}, whose number and the position relative to the Fermi energy are strongly influenced by the details of the magnetic order. It has been established that the value of the magnetic moment on Co ions is directly linked to the separation of Weyl nodes in the \textit{k}-space~\cite{Wang2018} while controlling the direction of magnetization allows to shift, create and annihilate Weyl nodes~\cite{Ghimire2019}.

The FM order is associated with itinerant electrons originating from \textit{d}-orbitals of Co ions ($\sim 0.3 \mu_\mathrm{B}$/Co), arranged in kagome layers (sketched in Figure 1a) and stacked along the \textit{c}-axis in the A-B-C pattern~\cite{Schnelle2013}. Such a reduced value of the magnetic moment has recently been explained within the local triangular Co cluster, giving rise to a $S = 1/2$ state ($ = 1 \mu_\mathrm{B}$) over three Co ions. It has been indicated numerically~\cite{Xu2018} and experimentally~\cite{Shen2019} that the moment is aligned parallel to the \textit{c}-axis, with a field of 23\,T needed to fully orient the moment parallel to the kagome plane~\cite{Shen2019}.

This simple configuration has been recently challenged by a muon-spin rotation ($\mu$SR) study, indicating that just below $T_\mathrm{C}$ the FM order is accompanied by a second phase, interpreted as an antiferromagnetic (AFM) arrangement of in-plane moments forming a type of a 120$^\circ$ configuration~\cite{Guguchia2020}. The second phase appears above 90\,K, reaching almost 70\% of the volume fraction just below $T_\mathrm{C}$. Those conclusions have been called into question by a recent polarized neutron study~\cite{Soh2021}, severely limiting the extent of a \textit{coherent} AFM phase. Additionally, a second transition has been suggested to appear in the range 125 -- 130\,K based on DC and AC magnetization measurements~\cite{Kassem2017}. Further experimental evidence for its presence has been accumulated by resistivity and magnetization~\cite{Lachman2020}, magneto-optical Kerr effect studies~\cite{Lee2021} and neutron diffraction~\cite{Soh2021}, although the exact temperature of the transition has been shown to differ from one study to the other. The observation of shifted, magnetic-field-driven hystereses at low temperatures has been ascribed to exchange bias mechanism and associated with the appearance of a glassy state below 125\,K~\cite{Lachman2020}.

In this Letter, we aim to clarify several confusing, and often contradicting findings related to the magnetic order in \css. We clearly establish strongly direction-dependent magnetic response and reveal that for $B \parallel c$ it is largely dominated by the demagnetization factor. For $B \perp c$ a second transition is found at $T_\mathrm{P} = 128$\,K, with the magnetic component strictly confined to the kagome plane, indicating small canting of magnetic moments on Co ions. We confirm the canting scenario by performing careful density-functional theory (DFT) calculations, revealing the umbrella structure where moments on a triangle tilt towards its center.

High-quality single crystals were grown by the self-flux method and characterized in previous publications~\cite{Liu2019,Xu2020,Soh2021}. DC and AC magnetization were measured on a SQUID-based magnetometer (MPMS3, Quantum Design) equipped with a horizontal rotator, allowing the angular precision better than 0.1$^\circ$. Three samples have been used throughout this study, cut from the same large single crystal, their dimensions and masses are detailed in Supplementary Material~\cite{SM}. DFT calculations were performed using the Vienna {\it Ab Initio} Simulation Package (VASP)\,\cite{Kresse96,Kresse93} on periodic models of the experimental crystal structure\,\cite{Vaqueiro09}. Electron-core interactions were described with the projector-augmented wave method, while the Kohn-Sham wave-functions for the valence electrons were expanded in a plane wave basis with a cutoff of 500 eV on the kinetic energy. We used a mesh of 11 $\times$ 11 $\times$ 11 $k$-points and the PBE variant\,\cite{PBE96} of the GGA functional while setting the convergence limit of energy to $10^{-8}$ eV. SOC was included in all calculations in a self-consistent manner.


\begin{figure}[t]
\centering
\includegraphics[width=0.55\columnwidth]{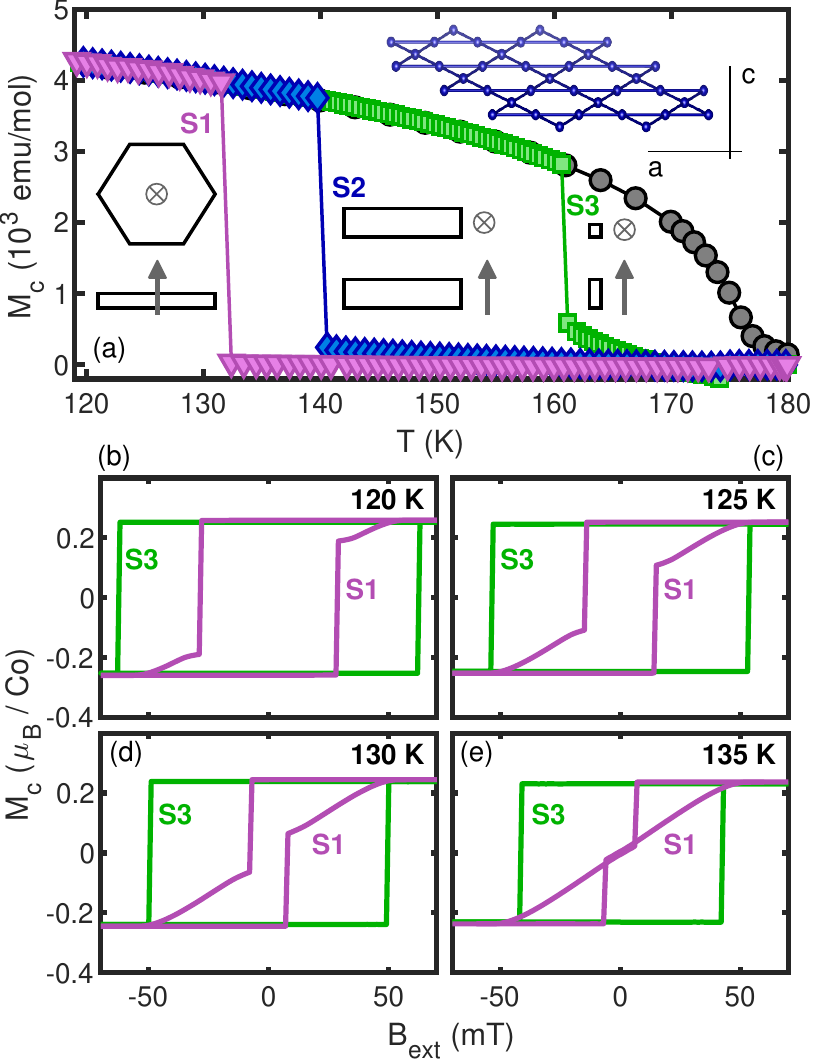}
\caption{(a) $M \parallel c$ vs temperature for samples $S1$, $S2$ and $S3$ following the FC/ZFW protocol. The samples were cooled down in $B_\mathrm{ext} = 10$\,mT (black circles). The arrows and crosses indicate the direction of the \textit{c}-axis. The panel also shows a single kagome plane of Co atoms. (b) - (e) Hystereses curves for samples $S1$ and $S3$.}
\label{fig:demagnetization}
\end{figure}

In several publications~\cite{Lachman2020,Lee2021,Soh2021} the appearance of the second transition has been reported as a jump in magnetization $M$ following a protocol in which the sample is cooled down to the lowest temperatures in magnetic field ($B \parallel c$) large enough to ensure the magnetization is completely saturated, after which $M(T)$ has been measured while warming in a nominally zero field (FC/ZFW). In Figure 1a we present our results following the same protocol for three different samples, with three markedly different temperatures at which the jump occurs. The observed variation can be associated with shapes of the prepared samples (sketched alongside the respective curves) and their demagnetization factors $N$~\cite{Chen1991,Chen2002}. During a ZFW protocol, although $B_{ext} = 0$, a fully saturated sample experiences a demagnetization field $H_d = -NM$ whose effect is to cause a reversal at a lower temperature when $N$ is larger. In the same way one can shift the reversal to lower or higher temperatures if $B_{ext} < 0$ or $B_{ext} > 0$, respectively~\cite{SM}. Therefore, the jumps in magnetization $M \parallel c$ during the FC/ZFW protocols cannot be taken as an evidence of a thermodynamic transition occurring around 125\,K while their observation in that range of temperatures is a simple consequence of a specific sample geometry, typical for thin samples of~\css. Additional evidence can be found in Figure 1b-e where hystereses loops are presented for $S1$ and $S3$ samples in the range of temperatures spanning the purported transition. In Ref.~\cite{Lachman2020} it has been suggested that one of the characteristic features is the change from the square-shaped loop into a more elongated one, with triangular wings. To the contrary, our results clearly show that neither $S1$ (large $N$) nor $S3$ (small $N$) change their behavior across 125\,K. $S1$ exhibits a square-loop only below 115\,K while $S3$ up to 155\,K. The only qualitative change occurs between 130\,K and 135\,K where the reversal field for $S1$ goes from negative to positive, which agrees with the observed reversal temperature in the ZFW protocol of 132\,K. Together with a lack of any feature seen in the AC response within the same temperature range (see below), we argue that there is no thermodynamic transition associated with $B \parallel c$ below $T_C$.


\begin{figure}[t]
\centering
\includegraphics[width=0.55\columnwidth]{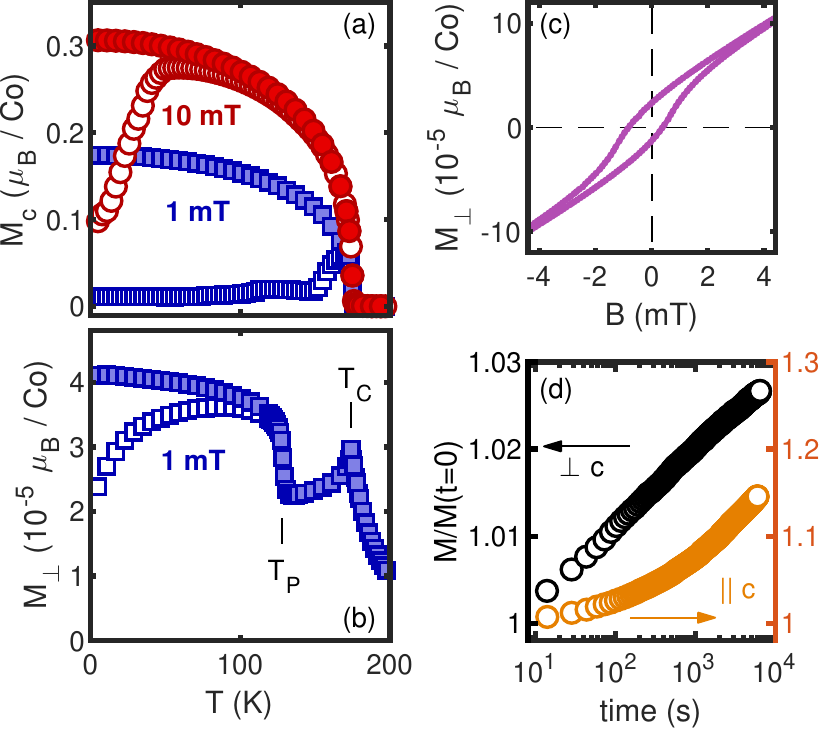}
\caption{(a) ZFC/FC protocols with $B \parallel c$ on $S3$. (b) ZFC/FC protocol with $B \perp c$ in 1\,mT on $S1$. (c) The hysteresis loop at $T = 15$\,K ($B \perp c$). (d) Time dependence of magnetization after ZFC to 50 K and $B = 1$\,mT for two directions.}
\label{fig:ZFCFC}
\end{figure}

On the other hand, evidence presented by Kassem \textit{et al.}~\cite{Kassem2017} and especially recent reports~\cite{Lachman2020,Zhang2021} for $B \perp c$ unequivocally point to an intrinsic feature related to the magnetism of \css. In order to directly compare contributions to magnetization along different crystallographic directions we plot in Figure 2 $M(T)$ for $B \parallel c$ and $B \perp c$ following standard zero-field cooled (ZFC)/field-cooled (FC) protocols. The main transition at $T_\mathrm{C}$ causes a sharp increase in $M_c$, followed by a ZFC/FC splitting stemming from the appearance of domains and pinning of domain walls. At the same time a clear maximum is observed in $M_{\perp}$ at $T_\mathrm{C}$. On the other hand, a clear indication of a second transition is seen only in $M_{\perp}$. In a similar fashion as $M_c$ increases and splits into separate ZFC and FC branches at $T_\mathrm{C}$, $M_{\perp}$ follows the same behavior below $T_\mathrm{P} = 128$\,K, with values approximately four orders of magnitude smaller than $M_c$. The associated hysteresis loop, presented in Figure 2c, reveals that the splitting collapses already at 2\,mT, significantly lower than for $B \parallel c$. An unequivocal decoupling of two directions can also be seen following a ZFC protocol to $T < T_P$ and observing a time dependence of magnetization after magnetic field is turned on. Figure 2d reveals that the in-plane magnetization process cannot be seen as a simple projection of a (much larger) signal in $M_c$ but it reflects a separate physical mechanism.


\begin{figure}[t]
\centering
\includegraphics[width=0.55\columnwidth]{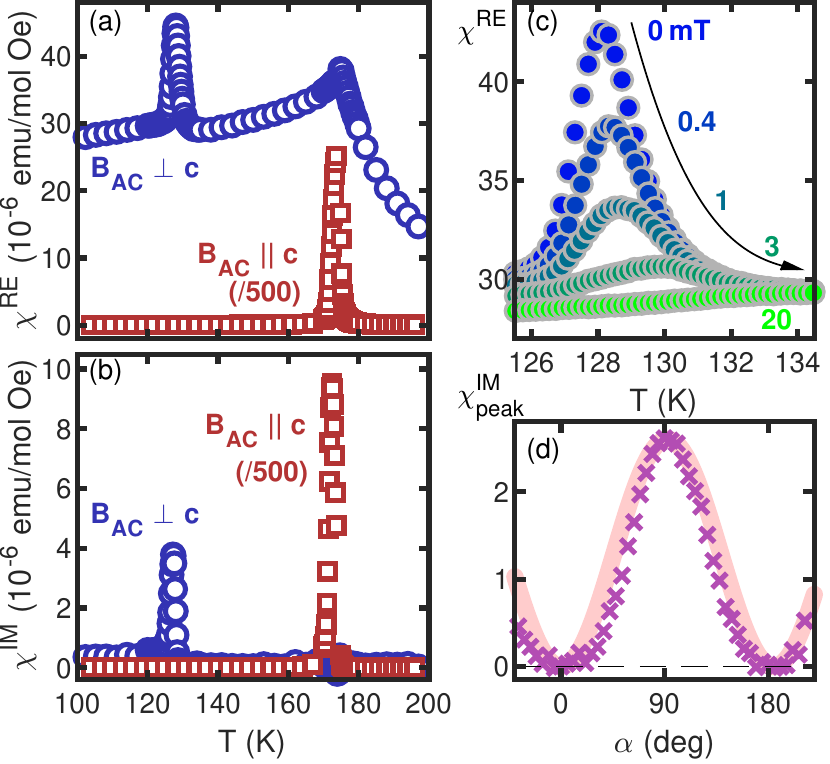}
\caption{(a) Real and (b) imaginary component of AC susceptibility ($B_{AC} = 1$ Oe) in $B = 0$ measured with 9.1\,Hz on $S2$. (c) $B$-dependence of $\chi^{RE}$ around $T_\mathrm{P}$ measured with 911\,Hz. The amplitude of the peak decreases with increasing $B$, as indicated by the arrow. (d) The angular dependence of the peak in $\chi^{IM}$ at $T_\mathrm{P}$ measured with 911\,Hz (crosses). The solid line indicates the $\sin^2{\alpha}$ dependence (see text).}
\label{fig:ACsusc}
\end{figure}

In Figures 3a and 3b we demonstrate the overall behavior of AC susceptibility, the real $\chi^{RE}$ and the imaginary component $\chi^{IM}$, for two directions in zero DC magnetic field. For $B_{AC} \parallel c$ both $\chi^{RE}$ and $\chi^{IM}$ show a sharp peak at $T_\mathrm{C}$ but no visible trace of the second transition. With $B_{AC} \perp c$ $\chi^{RE}$ shows a maximum at $T_\mathrm{C}$, similar in shape to $M_{\perp}$, without any sizeable feature in $\chi^{IM}$. At $T_\mathrm{P}$ both components reveal a sharp, narrow peak, below which $\chi^{RE}$ continues a slow decrease while $\chi^{IM}$ remains near zero. Again the scale of response is orders of magnitude smaller for $B_{AC} \perp c$ but the features all remain well defined. There is a very weak frequency dependence of the maximum at $T_\mathrm{P}$, similar in magnitude to the previous report~\cite{Lachman2020}.

One of the factors that strongly impacts the magnitude of the response at $T_\mathrm{P}$ is DC magnetic field. As seen in Figure 3c, the peak is more than halved at 1\,mT, while practically completely vanishes at 20\,mT. One should note that a superconducting magnet, after being ramped down to a nominal zero field, still contains stray fields from pinned vortices, which can reach several mT (and cause a negative signal in nominally ZFC measurement conditions~\cite{Liu2018,Kassem2017}). Combined with a typical sample mass of a couple of milligrams, it is easy to understand why in most cases the transition is hard to observe.

In order to reveal the angular dependence of the magnetic response around $T_\mathrm{P}$, we turn to $\chi^{IM}$. As displayed in Figure 3d, the amplitude of the peak closely follows a $\sim \sin^2{\alpha}$ dependence, where $\alpha$ is the angle between the AC magnetic field and the \textit{c}-axis. Such a type of a behavior is typical for projection-based vectors: one $\sin{\alpha}$ comes from the projection of the magnetic field onto a given axis, the second one comes from the projection of magnetic moments back to the axis of the magnetic field. This indicates that the dissipation induced around $T_\mathrm{P}$, and thus the order developing below $T_\mathrm{P}$ revealed by $M_{\perp}$ in Figure 2b, comes from the component of magnetic moments strictly perpendicular to the \textit{c}-axis. Note that $\chi^{IM}$ completely vanishes for $\alpha = 0$, unlike $\chi^{RE}$ which remains non-zero in the whole temperature range below $T_\mathrm{C}$ (Figure 3a). The lack of the imaginary component is a direct evidence against the idea of a thermodynamic transition occurring at $T_P$ since crossing from the low-temperature phase into a ferromagnetic phase above $T_P$ would involve a reconfiguration of moments and a substantial, more isotropic $\chi^{IM}$.

Such a strictly in-plane component is incompatible with any known spin-glass scenario, which has been suggested solely based on a weak frequency dependence at $T_P$. It is left unclear how this state would emerge from an FM state with a strong spin anisotropy or where is disorder needed for a glassy state coming from. On the other hand, the weak in-plane dynamics can arise if the dominantly \textit{c}-axis oriented magnetic moments are canted by a small angle. The question of canting of the moments in \css\,has been addressed in several publications. Xu \textit{et al.}~\cite{Xu2018} were first to use DFT calculations to calculate the total energy vs the canting angle, suggesting $M \parallel c$. Ghimire \textit{et al.}~\cite{Ghimire2019} studied the magnetic field-induced canting and its effect on the position and the total number of Weyl nodes. This was followed up by the $\mu$SR study~\cite{Guguchia2020} which suggested a sizeable in-plane component forming a separate AFM phase, coexisting with the main (FM) one. Finally, recent DFT calculations by Solovyev \textit{et al.}~\cite{Solovyev2022} addressed the stability of the FM order and again concluded that the lowest energy configuration corresponds to moments being strictly parallel to the \textit{c}-axis. Contrary to those results is the more general, symmetry-based discussion about complex noncollinear structures in itinerant magnets with non-negligible SOC~\cite{Sandratskii1998}, which helped explain the emergence of canting in several systems, including Mn$_3$Sn and U$_3$P$_4$. The main conclusion is that the ferromagnetic structure with $M \parallel c$ is not symmetry-protected, allowing for SOC-induced, small-angle canting of moments.

Motivated by these considerations, we performed self-consistent non-collinear magnetic calculations with VASP. We find that the magnetic moments in the resulting ferromagnetic ground state deviate from the $c$-axis by a rather small angle $\theta_0 \approx 1.5^\circ$ forming the ``umbrella structure'' sketched in Figure 4a. The magnetic moment vector obtained from the self-consistent calculations for the three Co sites are $(-0.008,-0.004,0.345)$, $(0.008,-0.004,0.345)$ and $(0,0.009,0.345)$ in Bohr magneton units. We find that the in-plane components of the moments are quite small, but very robust to the changes in $k$-space mesh and initial electron density guess. Additionally, we calculated the total energy of the system with respect to the canting angle $\theta$ (focusing especially on the small-angle region), by constraining the direction of the moments at the three Co sites while allowing the size of the moments to evolve self-consistently. As shown in Figure 4b, the results of these calculations confirm that the deviation of moments from the \textit{c}-axis by an angle of $\theta_0 \approx 1.5^\circ$, yields lower total energy per Co ion. 
It should be remarked that both previous reports~\cite{Xu2018,Solovyev2022} considered a relatively sparse grid, with steps in the range 5 -- 10$^\circ$. Additionally, Solovyev \textit{et al.}~\cite{Solovyev2022} performed the calculations without SOC, therefore our results represent the first detailed numerical investigation of the full SOC-driven magnetic structure in \css. Note that the umbrella structure is fully compatible with a single $\Gamma^{+}_{2}$ irreducible representation~\cite{Soh2021}.


\begin{figure}[t]
\centering
\includegraphics[width=0.55\columnwidth]{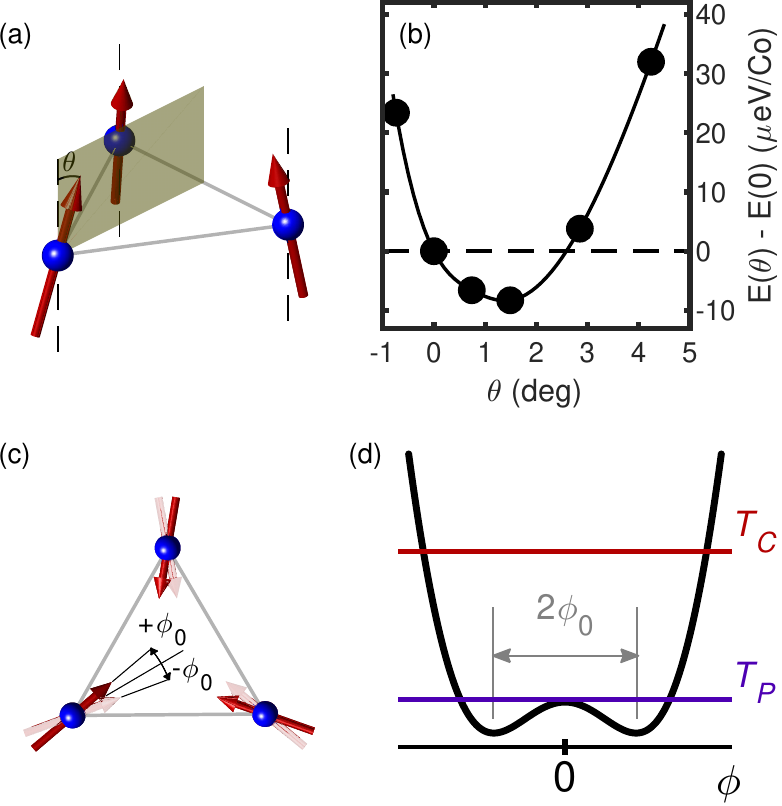}
\caption{(a) Sketch of the umbrella structure for $\theta > 0^\circ$. The moments rotate in the plane passing through the center of the triangle (dark yellow). (b) Total energy per cobalt ion for different angles relative to the energy at $\theta = 0^\circ$, as calculated by constrained DFT calculations. (c) The top view of the triangle with in-plane projections having two possible orientations given by $+\phi_0$ and $-\phi_0$. (d) The sketch of the energy potential with respect to the angle $\phi$. Due to the small size of $\phi_0$, the energy barrier between two orientations traps the moments only at $T = T_\mathrm{P} < T_\mathrm{C}$.}
\label{fig:DFT}
\end{figure}

We can return now to the question of the second transition at $T_\mathrm{P}$ and the development of the FM-like behavior of $M_{\perp}$ below $T_\mathrm{P}$. It is evident that the FM component cannot arise from the umbrella structure as presented in Figure 4a, since the spin projections onto the kagome plane are forming a 120$^\circ$ structure, which is a fully compensated AFM arrangement. Instead, we propose a further modification which involves the rotation of moments around the \textit{c}-axis by an angle $\phi_0$, Figure 4c. Given that the transition at $T_C$ is described within $\Gamma^{+}_{2}$~\cite{Soh2021}, such a rotation is forbidden by symmetry. However, once the temperature is lowered below $T_C$ and the ordered state has been established, there are no symmetry-related constraints imposed on this type of a rotation. A simple model, presented in Figure 4d, encapsulates both of these aspects. It consists of a double-well potential with a barrier height significantly lower than $T_C$. Because the potential is symmetric in $\phi$, the fluctuations of the order parameter within the critical region around $T_C$ are still described by $\Gamma^{+}_{2}$ since $<\phi> = 0$. Once the temperature is lowered to $T_P$, the $\phi$-rotations start to randomly freeze into $+\phi_0$ and $-\phi_0$, giving rise to the peak in AC susceptibility and its weak frequency dependence. A very small magnetic field can switch and eventually completely polarize those projections, ultimately giving rise to the dynamic completely different from the one observed for domain walls for $B || c$. Alternatively, the double-well potential could arise \textit{at} $T_P$, implying an additional energy scale for which there is no immediate evidence.

The value of $\phi_0$ can be estimated from the in-plane FM component which is given by $M_{\perp} \sim M_c \sin{\theta_0}\sin{\phi_0}$. Taking that $M_{\perp}/M_c \sim 10^{-4}$ and $\theta_0 \approx 1.5^\circ$, the amount of rotation perpendicular to the \textit{c}-axis is $\phi_0 \sim 0.2 - 0.3^\circ$. Such a small value of $\phi_0$ agrees with the model's assumption of a very small energy barrier while on the other hand represents a significant challenge for DFT and can even arise from contributions not considered in our calculations. It also explains the lack of any feature in specific heat, since the entropy change across $T_P$ would be negligibly small compared to phonon and magnon contributions.

Our findings are based on a careful control of all experimental parameters using large, high-quality single crystals. As such, they pose severe limits to hypotheses about the coexistence of secondary phases~\cite{Guguchia2020,Lachman2020}. Note that those two proposals are mutually contradictory: the in-plane, AFM phase has been suggested to occur above $~\sim 90$ K~\cite{Guguchia2020}, reaching almost 70\% volume fraction close to $T = T_\mathrm{C}$ and having a different ordering temperature $T_\mathrm{AFM} < T_\mathrm{C}$. On the other hand, the spin-glass phase has been hypothesized to exist below $T = T_\mathrm{P}$. No indication of $T_{AFM}$ could be found below $T_\mathrm{C}$, nor were there any evidence of the dynamics of the phase boundary between the FM and AFM volume fractions. We confirmed the weak frequency dependence at $T_\mathrm{P}$ but for a true spin-glass phase the response would be expected to be more isotropic. We emphasize that many observations presented in those studies could find a natural explanation in a peculiar behavior of domain walls~\cite{Schnelle2013,Lee2021}.

In summary, our results resolve several outstanding issues in \css. The second transition occurs at $T_P = 128$\,K and is related to the freezing of a small component of magnetic moments perpendicular to the $c$-axis. The jumps observed in magnetization parallel to the $c$-axis are revealed to be sample dependent through the effect of demagnetization. Our SOC-based DFT calculations found that the moments are slightly canted away from the \textit{c}-axis, forming the umbrella configuration. The additional canting along the $\phi$-angle is needed to account for the in-plane component, whose origin at the moment remains unknown.

I. {\v Z}. acknowledges fruitful discussions with V. Ivanov, S. Savrasov, S. Acharya and M. I. Katsnelson. J.-R. Soh acknowledges support from the Singapore National Science Scholarship, Agency for Science Technology and Research. I. {\v Z}. and J.-R. Soh acknowledge support from the ERC Synergy grant HERO (Grant ID: 810451). C.Y. and Y.S. were supported by the National Natural Science Foundation of China (No. 12004416, No. U2032204) and the Informatization Plan of the Chinese Academy of Sciences (CAS-WX2021SF-0102). R.Y. was supported by the Swiss National Science Foundation (SNSF) Sinergia network `NanoSkyrmionics' (grant No. CRSII5-171003). H. M. R. acknowledges the support from SNSF Projects No. 200020-188648 and 206021-189644. First-principles calculations were performed at the Swiss National Supercomputing Centre (CSCS) under project s1008.

\bibliography{ref}

\clearpage

\section{SUPPLEMENTARY MATERIAL}

\section{Resistivity}

Resistivity was measured with a 4-point contact method on a well defined geometry for both directions: 3 x 0.9 x 0.6 mm$^3$ for $I \parallel ab$ and 2 x 1 x 0.5 mm$^3$ for $I \parallel c$.

\begin{figure*}[h]
    \centering
    \includegraphics[width=0.5\columnwidth]{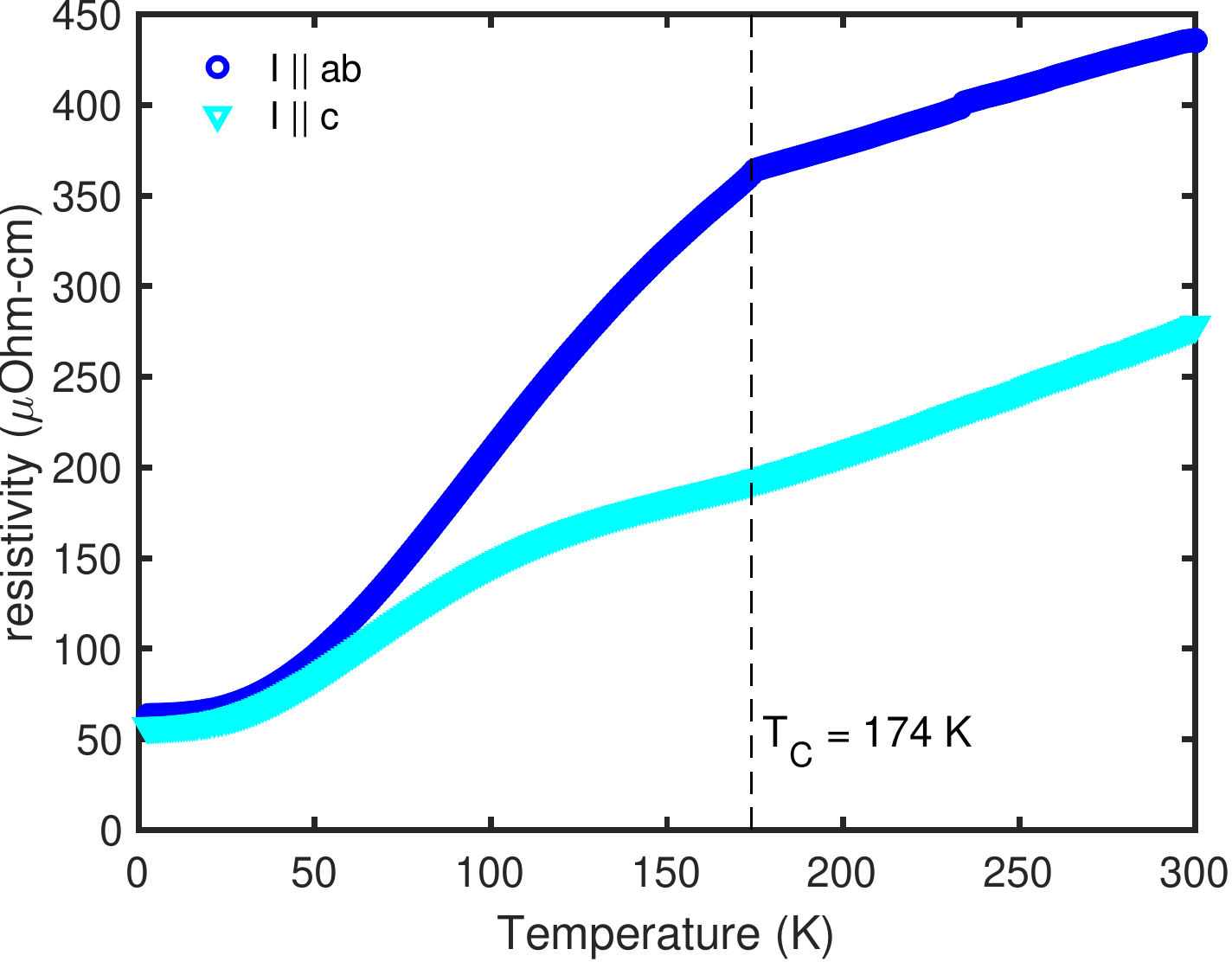}
    \caption{Resistivity of \css.}
    \label{Sfig:resistivity}
\end{figure*}

\pagebreak

\section{Curie-Weiss behavior}

The magnetization of the sample has been measured in $B = 0.1$ T. A fit to the Curie-Weiss formula $\chi = C/(T - T_C) + \chi_0$ has been performed above the ordering temperature. The extracted values are $C = 0.3845$ emu K/mol, $T_C = 184$ K and $\chi_0 = -0.00017$ emu/mol. Here we have normalized the magnetization to the full formula unit (3 Co ions). The value of $S = 1/2$/(3 Co) should result in $C = 0.375$ with $g = 2$, in excellent agreement with our measurements.

\begin{figure*}[h]
    \centering
    \includegraphics[width=0.5\columnwidth]{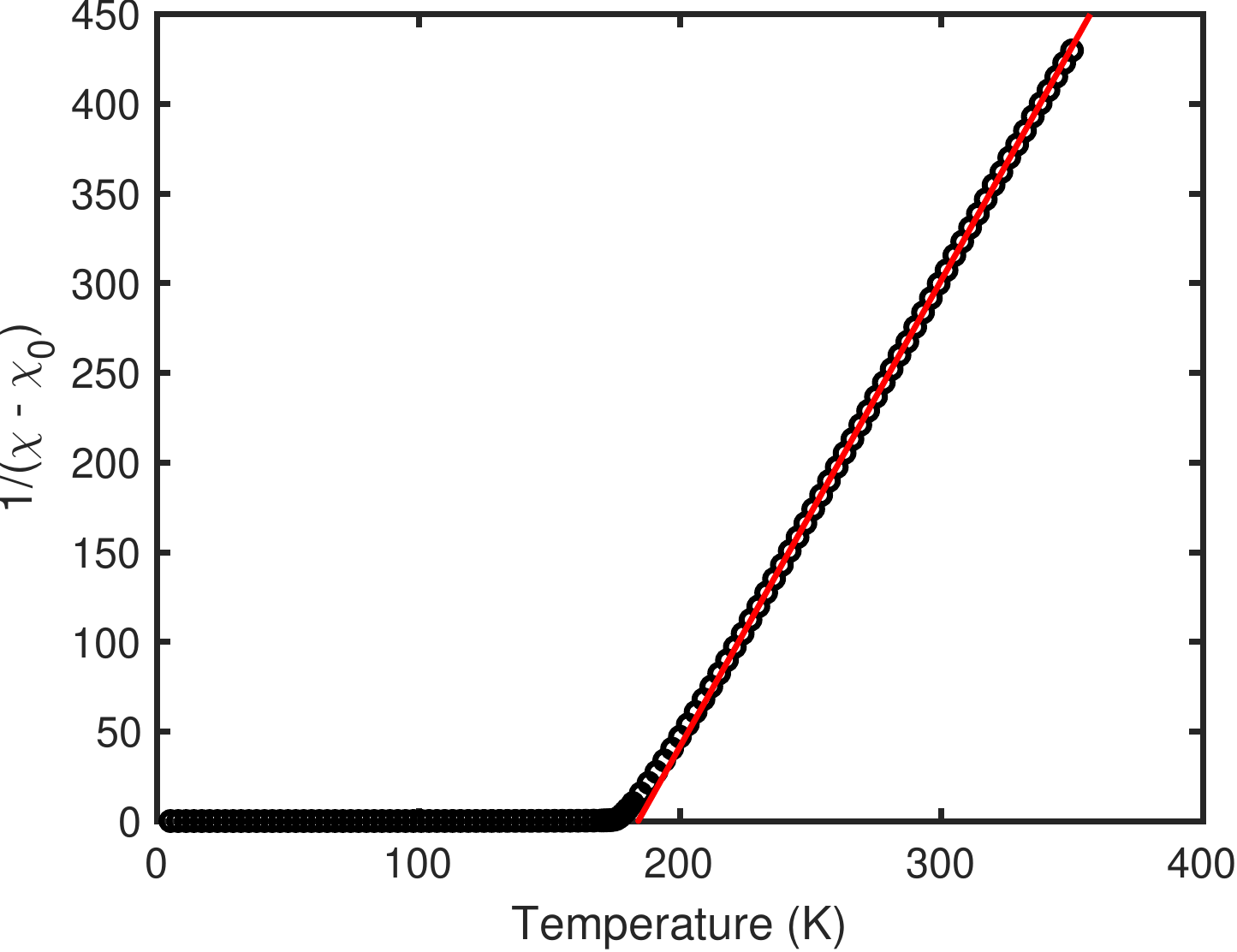}
    \caption{Temperature dependence of the inverse susceptibility. The red line represents the Curie-Weiss behavior.}
    \label{Sfig:CW}
\end{figure*}
%


%
\begin{table}
\begin{tabular}{lccc}
\hline
Sample & Dimensions (mm) & Mass (mg) & \textit{N} \\
\hline
$S1$ & 5 x 5 x 1 & 106.6 & $\sim 0.68$\\

$S2$ & 6 x 2 x 2 & 153.3 & $\sim 0.43$ \\

$S3$ & 0.5 x 0.5 x 2 & 3.23 & $\sim 0.11$ \\
\hline
\end{tabular}
\caption{Dimensions, masses and calculated demagnetization factors~\cite{Chen1991,Chen2002} for three samples of \css.}
\label{table:samples}
\end{table}

\pagebreak

\section{Magnetization reversal}

\begin{figure*}[h]
    \centering
    \includegraphics[width=0.5\columnwidth]{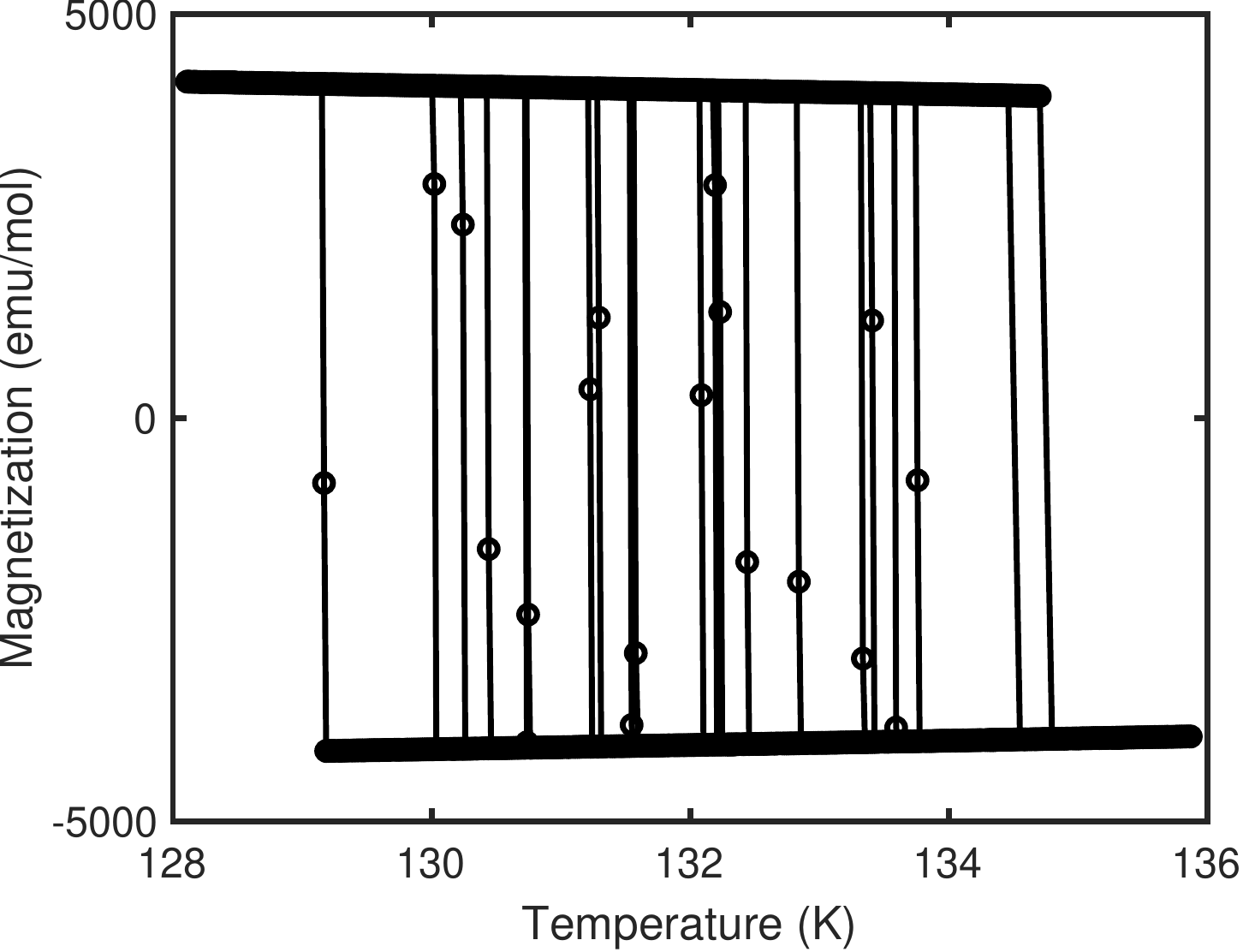}
    \caption{Reversals of magnetization for sample $S1$ with varying magnetic field, ranging from -6 mT to +6 mT.}
    \label{Sfig:S1reversals}
\end{figure*}
\begin{figure*}[h]
    \centering
    \includegraphics[width=0.5\columnwidth]{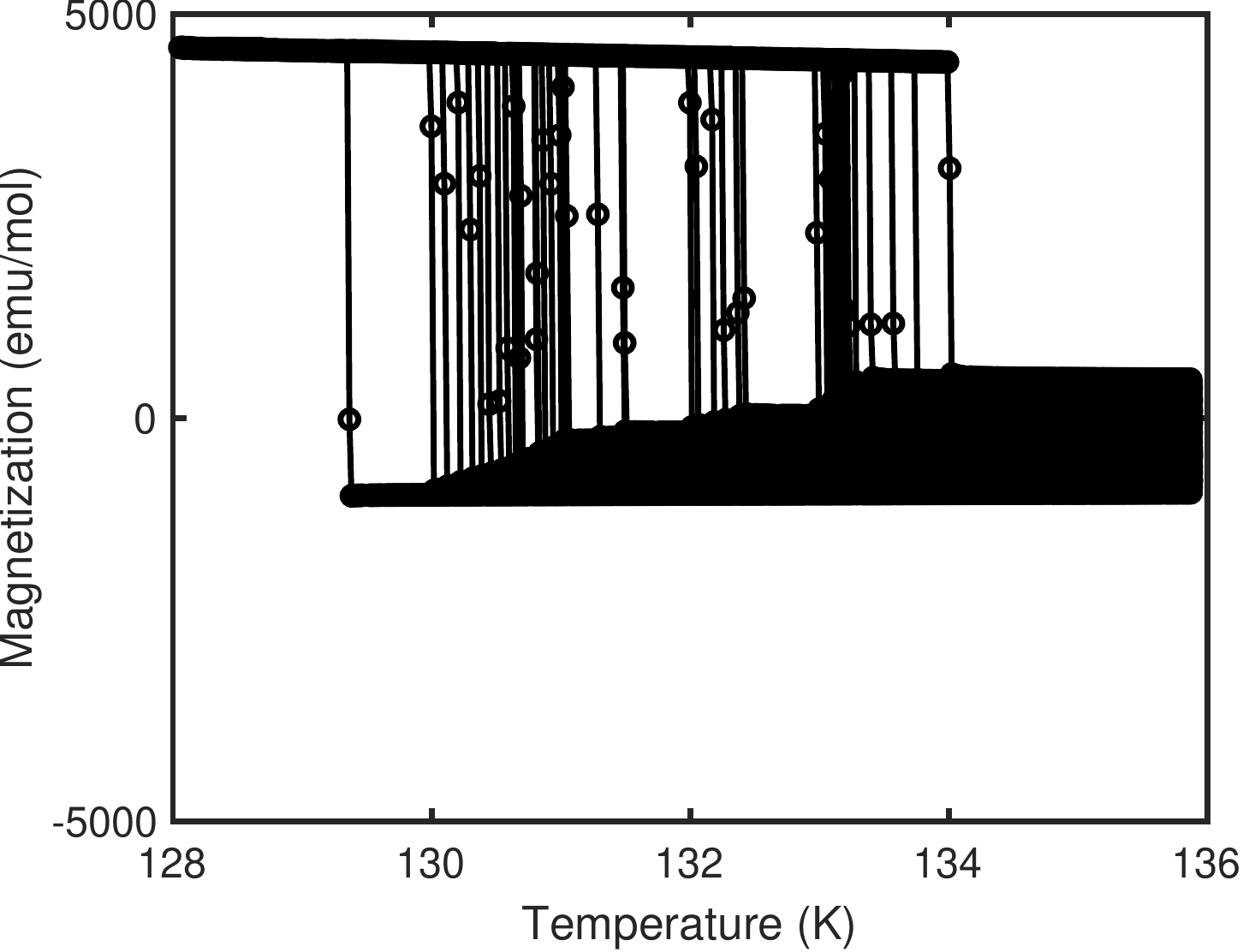}
    \caption{Reversals of magnetization for sample $S2$ with varying magnetic field, ranging from -4.5 mT to +1 mT, which causes the reversal to occur around 132 K.}
    \label{Sfig:S2reversals}
\end{figure*}
\begin{figure*}[h]
    \centering
    \includegraphics[width=0.5\columnwidth]{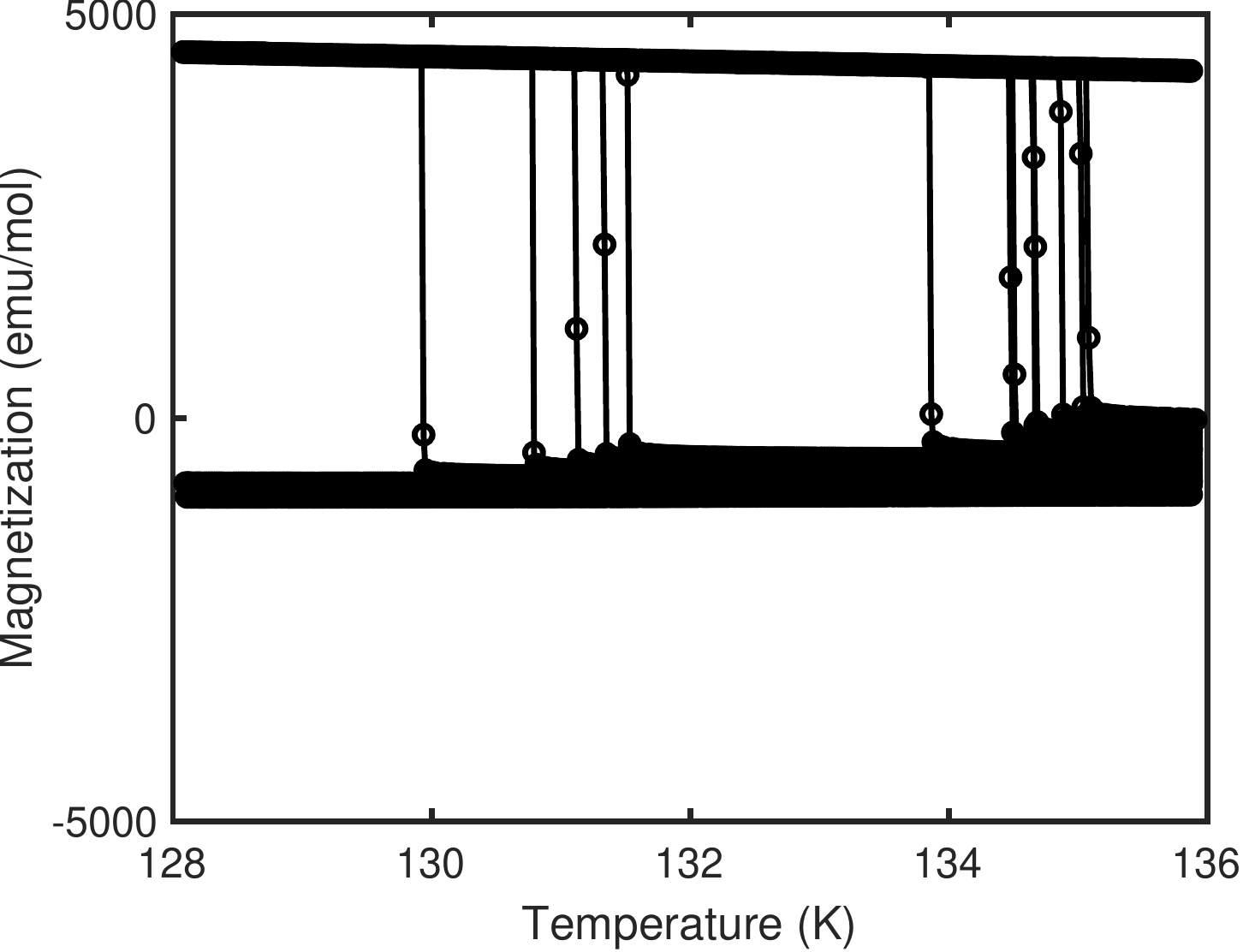}
    \caption{Reversals of magnetization for sample $S3$ with varying magnetic field, ranging from -52 mT to -45 mT, which causes the reversal to occur around 132 K.}
    \label{Sfig:S3reversals}
\end{figure*}
\begin{figure*}[h]
    \centering
    \includegraphics[width=0.5\columnwidth]{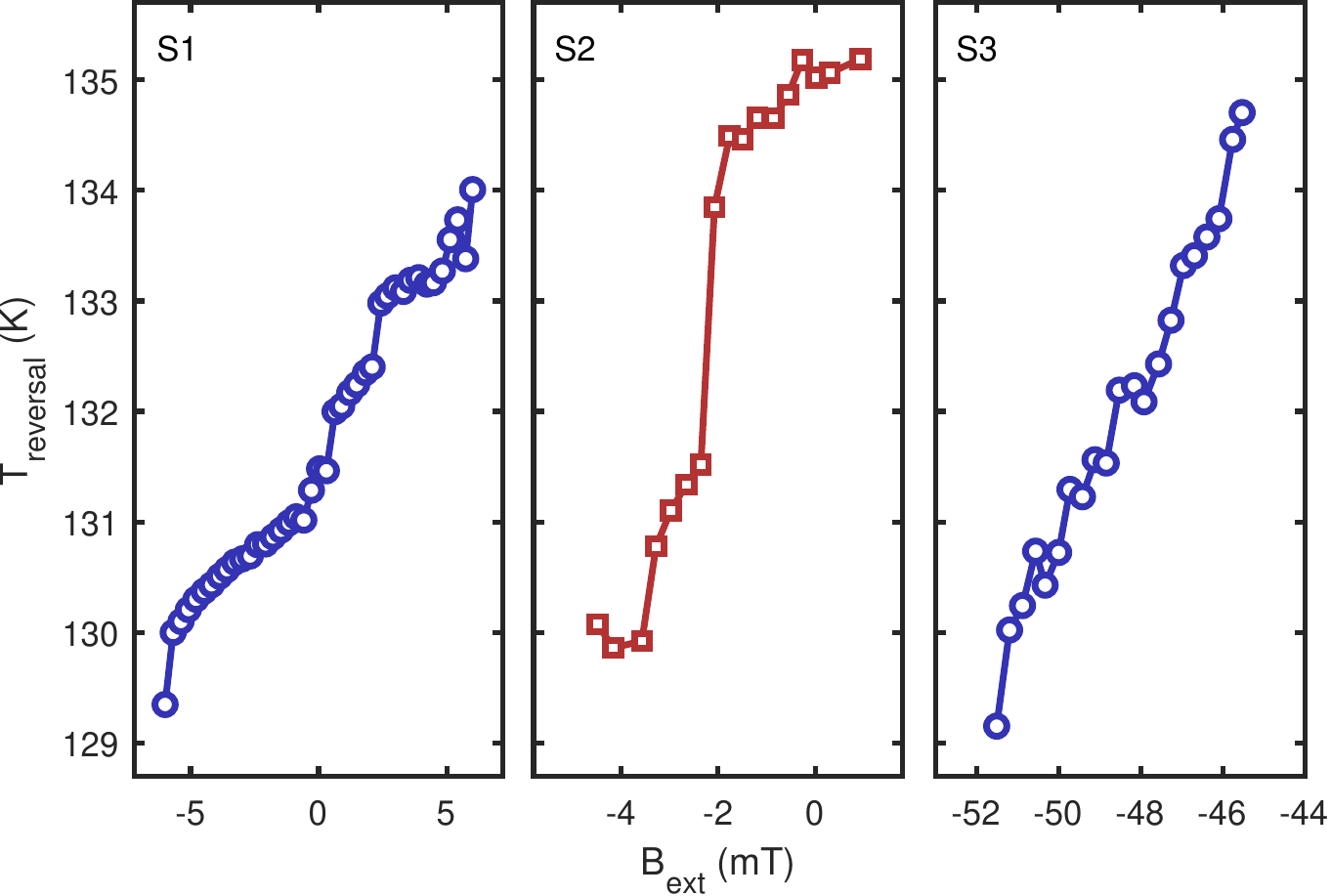}
    \caption{Temperature and magnetic field dependence of the reversal for three samples.}
    \label{Sfig:Treversal}
\end{figure*}

\clearpage

\section{Comparison between three samples}

\begin{figure*}[h]
    \centering
    \includegraphics[width=0.9\columnwidth]{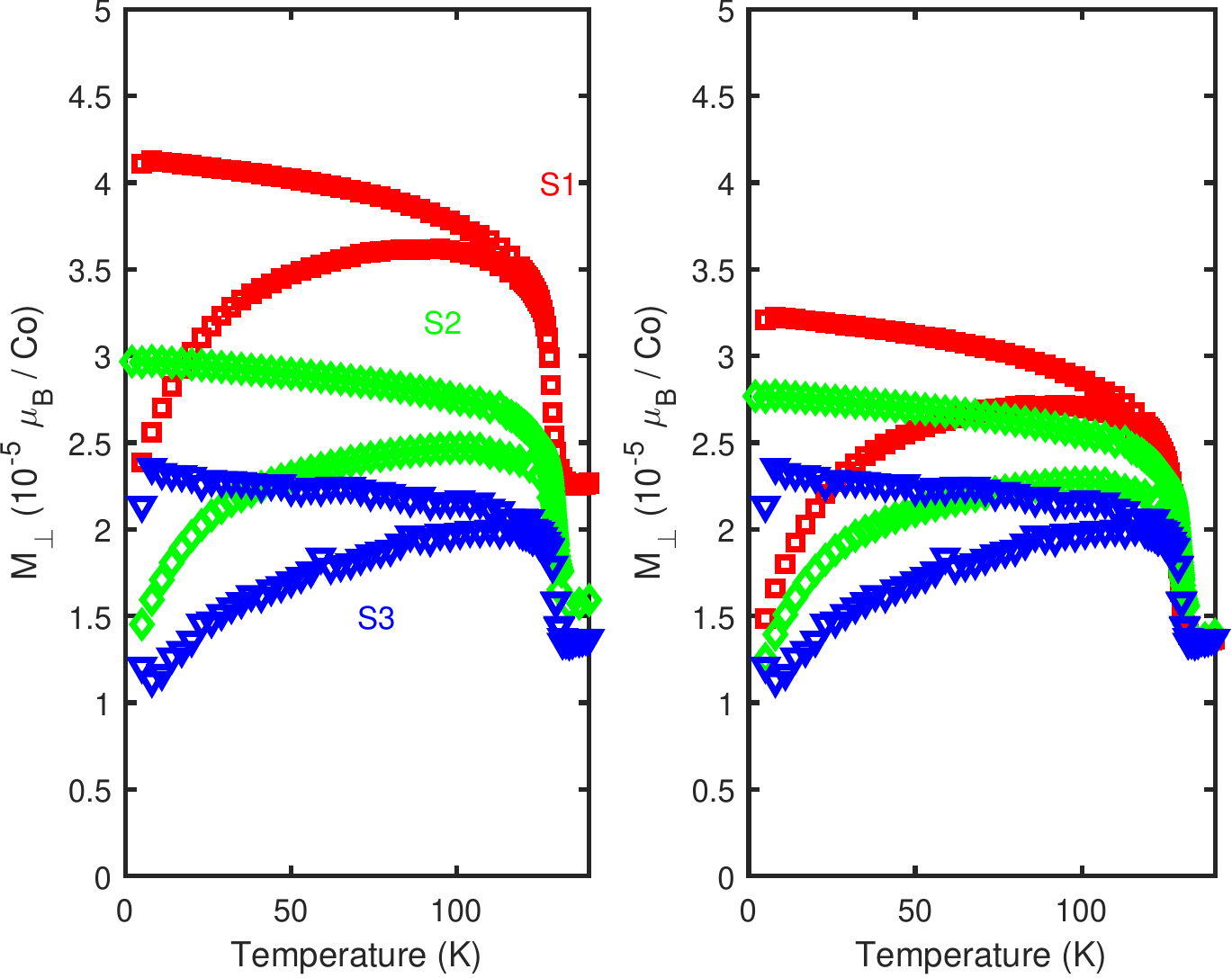}
    \caption{In-plane magnetization for three samples. (Left) as measured. (Right) all the curves overlapping above $T_P$. The observed variation is coming from difficulties while measuring weak in-plane signals with a large moment along the c-axis. Due to the inevitable presence of non-zero magnetic field while cooling (due to the remanence of the superconducting magnet or even due to the Earth's magnetic field) and \css being an extremely soft ferromagnet just below $T_C$, there is always a non-zero magnetization along the c-axis. Thus, when magnetic field is applied perpendicular to the c-axis, a small torque generated turns the sample together with the sample holder. This then can cause a fraction of the c-axis magnetization to contribute to the measured signal. We cannot exclude other potential contributions.}
    \label{Sfig:inplaneZFCFC}
\end{figure*}
\begin{figure*}
    \centering
    \includegraphics[width=0.5\columnwidth]{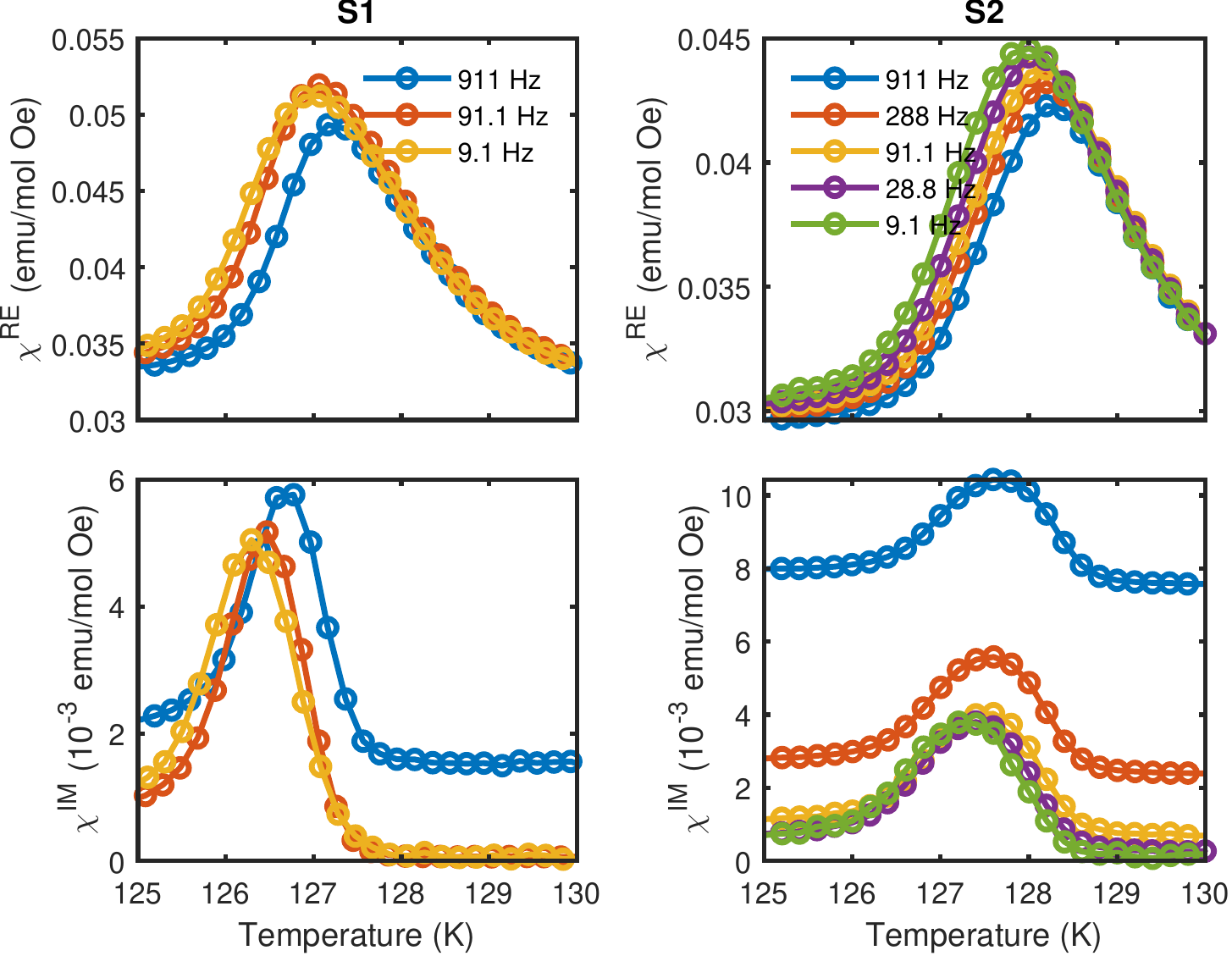}
    \caption{AC susceptibility for samples $S1$ and $S2$.}
    \label{Sfig:freqAC}
\end{figure*}

\pagebreak

\section{Angular dependence around the main transition}

\begin{figure*}[h]
    \centering
    \includegraphics[width=0.5\columnwidth]{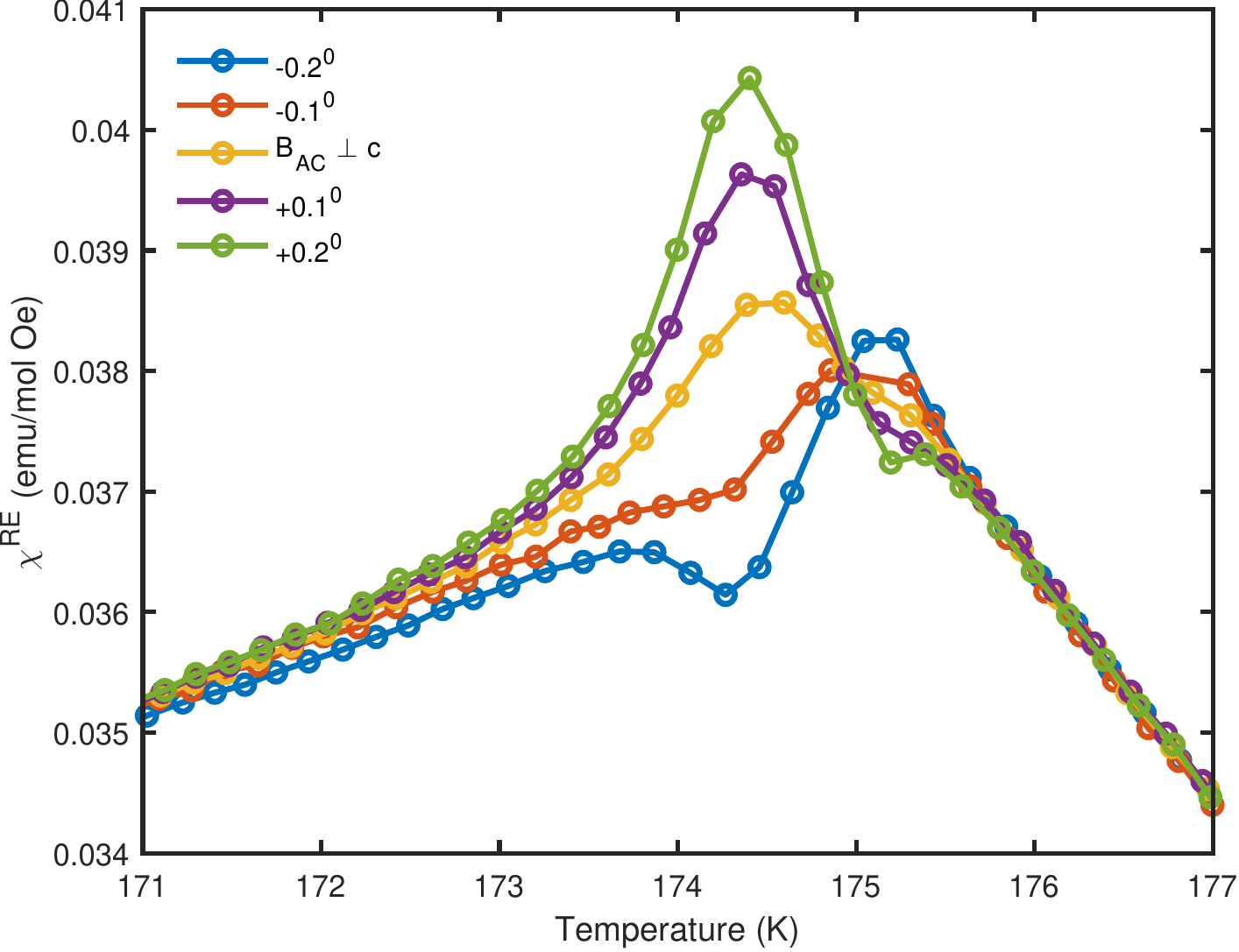}
    \caption{Angular dependence of the real component of the AC susceptibility around the main transition.}
    \label{Sfig:angleAC}
\end{figure*}

\end{document}